# Fano-enhanced pulling and pushing optical force on active plasmonic nanoparticles

D. L. GAO,[1,2] R. SHI,[1] Y. HUANG,[1] W. H. NI,[1] AND L. GAO[1,*]

[1]*College of Physics, Optoelectronics and Energy & Collaborative Innovation Center of Suzhou Nano Science and Technology, Soochow University, Suzhou 215006, China*
[2]*Department of Electrical and Computer Engineering, National University of Singapore, 4 Engineering Drive 3, Singapore 117576, Republic of Singapore*
*E-mail: leigao@suda.edu.cn

**Abstract:** We demonstrate tunable pulling and pushing optical forces on plasmonic nanostructures around Fano resonance. The plasmonic nanostructure containing a spherical core with optical gain and a metallic shell shows much larger optical pulling force than a pure gain sphere. One can obtain large field enhancement and giant pulling force at the emerged quadrupole mode. The introduction of optical pump compensate the dissipative loss from metal shell, thus enable the strong coupling between a narrow quadrupole mode and a board dipole mode, giving rise to Fano resonance. The giant negative forces origin from the reversal of electric field at Fano resonance, which lead to pulling force on bound currents and charges. Meanwhile, the separation of the Lorentz force helps to reveal the nature of the pulling forces in gain system. We have shown that by applying the Lorentz force density formula, it is possible to obtain the correct value of the force inside our complex inhomogenous structure made up of dispersive and lossy metamaterial irrespective of the electromagnetic momentum density. Our results provide a practical way to manipulate nanoparticles and give deep insight into light-matter interaction.

**OCIS codes:** (020.7010) Laser trapping; (290.4020) Mie theory; (290.5825) Scattering theory.

## References and links

1. J. J. Sáenz, "Optical forces: Laser tractor beams," Nat. Photonics **5**(9), 514-515 (2011).
2. S. Sukhov, and A. Dogariu, "Negative Nonconservative Forces: Optical "Tractor Beams" for Arbitrary Objects," Phys. Rev. Lett. **107**(20), 203602 (2011).
3. J. Chen, J. Ng, Z. Lin, and C. T. Chan, "Optical pulling force," Nat. Photonics **5**(9), 531-534 (2011).
4. A. Novitsky, C.-W. Qiu, and A. Lavrinenko, "Material-Independent and Size-Independent Tractor Beams for Dipole Objects," Phys. Rev. Lett. **109**(2), 023902 (2012).
5. A. Novitsky, C.-W. Qiu, and H. Wang, "Single Gradientless Light Beam Drags Particles as Tractor Beams," Phys. Rev. Lett. **107**(20), 203601 (2011).
6. J. M. Auñón, C. W. Qiu, and M. Nieto-Vesperinas, "Tailoring photonic forces on a magnetodielectric nanoparticle with a fluctuating optical source," Phys. Rev. A **88**(4), 043817 (2013).
7. A. Novitsky, and C.-W. Qiu, "Pulling extremely anisotropic lossy particles using light without intensity gradient," Phys. Rev. A **90**(5), 053815 (2014).
8. O. Brzobohatý, V. Karásek, M. Šiler, L. Chvátal, T. Čižmár, and P. Zemánek, "Experimental demonstration of optical transport, sorting and self-arrangement using a 'tractor beam'," Nat. Photonics **7**(2), 123-127 (2013).
9. S. H. Lee, Y. Roichman, and D. G. Grier, "Optical solenoid beams," Opt. Express **18**(7), 6988-6993 (2010).
10. Y. Roichman, B. Sun, Y. Roichman, J. Amato-Grill, and D. G. Grier, "Optical forces arising from phase gradients," Phys. Rev. Lett. **100**(1), 013602 (2008).
11. J. Curtis, and D. Grier, "Structure of Optical Vortices," Phys. Rev. Lett. **90**(13), 133901 (2003).
12. J. Ng, Z. Lin, and C. T. Chan, "Theory of Optical Trapping by an Optical Vortex Beam," Phys. Rev. Lett. **104**(10), 103601 (2010).
13. S. Albaladejo, M. I. Marques, M. Laroche, and J. J. Saenz, "Scattering forces from the curl of the spin angular momentum of a light field," Phys. Rev. Lett. **102**(11), 113602 (2009).
14. Y. Zhao, A. A. E. Saleh, and J. A. Dionne, "Enantioselective Optical Trapping of Chiral Nanoparticles with Plasmonic Tweezers," ACS Photonics **3**(3), 304-309 (2016).
15. L. Jauffred, S. M. Taheri, R. Schmitt, H. Linke, and L. B. Oddershede, "Optical Trapping of Gold Nanoparticles in Air," Nano Lett. **15**(7), 4713-4719 (2015).

16. M. Righini, P. Ghenuche, S. Cherukulappurath, V. Myroshnychenko, F. J. Garcia de Abajo, and R. Quidant, "Nano-optical trapping of Rayleigh particles and Escherichia coli bacteria with resonant optical antennas," Nano Lett. **9**(10), 3387-3391 (2009).
17. A. Mizrahi, and Y. Fainman, "Negative radiation pressure on gain medium structures," Opt. Lett. **35**(20), 3405-3407 (2010).
18. K. J. Webb, and Shivanand, "Negative electromagnetic plane-wave force in gain media," Phys. Rev. E **84**(5), 057602 (2011).
19. M.-T. Wei, and A. Chiou, "Three-dimensional tracking of Brownian motion of a particle trapped in optical tweezers with a pair of orthogonal tracking beams and the determination of the associated optical force constants," Opt. Express **13**(15), 5798 (2005).
20. H. Xu, and M. Käll, "Surface-Plasmon-Enhanced Optical Forces in Silver Nanoaggregates," Phys. Rev. Lett. **89**(24), 246802 (2002).
21. A. S. Zelenina, R. Quidant, and M. Nieto-Vesperinas, "Enhanced optical forces between coupled resonant metal nanoparticles," Opt. Lett. **32**(9), 1156 (2007).
22. L. Zhang, X. Dou, C. Min, Y. Zhang, L. Du, Z. Xie, J. Shen, Y. Zeng, and X. Yuan, "In-plane trapping and manipulation of ZnO nanowires by a hybrid plasmonic field," Nanoscale **8**(18), 9756-9763 (2016).
23. C. Yang, D. Pan, L. Tong, and H. Xu, "Guided transport of nanoparticles by plasmonic nanowires," Nanoscale **8**(46), 19195-19199 (2016).
24. H. Chen, L. Shao, Y. C. Man, C. Zhao, J. Wang, and B. Yang, "Fano resonance in (gold core)-(dielectric shell) nanostructures without symmetry breaking," Small **8**(10), 1503-1509 (2012).
25. O. Pena-Rodriguez, and U. Pal, "Au@Ag core-shell nanoparticles: efficient all-plasmonic Fano-resonance generators," Nanoscale **3**(9), 3609-3612 (2011).
26. O. Pena-Rodriguez, A. Rivera, M. Campoy-Quiles, and U. Pal, "Tunable Fano resonance in symmetric multilayered gold nanoshells," Nanoscale **5**(1), 209-216 (2013).
27. J. B. Lassiter, H. Sobhani, M. W. Knight, W. S. Mielczarek, P. Nordlander, and N. J. Halas, "Designing and deconstructing the Fano lineshape in plasmonic nanoclusters," Nano Lett. **12**(2), 1058-1062 (2012).
28. J. Zhang, and A. Zayats, "Multiple Fano resonances in single-layer nonconcentric core-shell nanostructures," Opt. Express **21**(7), 8426-8436 (2013).
29. B. Luk'yanchuk, N. I. Zheludev, S. A. Maier, N. J. Halas, P. Nordlander, H. Giessen, and C. T. Chong, "The Fano resonance in plasmonic nanostructures and metamaterials," Nat. Mater. **9**(9), 707-715 (2010).
30. Y. Yang, W. Wang, A. Boulesbaa, Kravchenko, II, D. P. Briggs, A. Puretzky, D. Geohegan, and J. Valentine, "Nonlinear Fano-Resonant Dielectric Metasurfaces," Nano Lett. **15**(11), 7388-7393 (2015).
31. Z. Li, S. Zhang, L. Tong, P. Wang, B. Dong, and H. Xu, "Ultrasensitive size-selection of plasmonic nanoparticles by Fano interference optical force," ACS Nano **8**(1), 701-708 (2014).
32. H. Chen, S. Liu, J. Zi, and Z. Lin, "Fano Resonance-Induced Negative Optical Scattering Force on Plasmonic Nanoparticles," ACS Nano **9**(2), 1926-1935 (2015).
33. T. Cao, L. Mao, D. Gao, W. Ding, and C.-W. Qiu, "Fano resonant Ge2Sb2Te5 nanoparticles realize switchable lateral optical force," Nanoscale **8**(10), 5657-5666 (2016).
34. J. Pan, Z. Chen, J. Chen, P. Zhan, C. J. Tang, and Z. L. Wang, "Low-threshold plasmonic lasing based on high-Q dipole void mode in a metallic nanoshell," Opt. Lett. **37**(7), 1181-1183 (2012).
35. N. K. Grady, N. J. Halas, and P. Nordlander, "Influence of dielectric function properties on the optical response of plasmon resonant metallic nanoparticles," Chem. Phys. Lett. **399**(1-3), 167-171 (2004).
36. G. Strangi, A. De Luca, S. Ravaine, M. Ferrie, and R. Bartolino, "Gain induced optical transparency in metamaterials," Appl. Phys. Lett. **98**(25), 251912 (2011).
37. S. Campione, M. Albani, and F. Capolino, "Complex modes and near-zero permittivity in 3D arrays of plasmonic nanoshells: loss compensation using gain [Invited]," Opt. Mater. Express **1**(6), 1077 (2011).
38. A. Fang, Z. Huang, T. Koschny, and C. M. Soukoulis, "Overcoming the losses of a split ring resonator array with gain," Opt. Express **19**(13), 12688-12699 (2011).
39. E. Prodan, C. Radloff, N. J. Halas, and P. Nordlander, "A hybridization model for the plasmon response of complex nanostructures," Science **302**(5644), 419-422 (2003).
40. E. Prodan, and P. Nordlander, "Plasmon hybridization in spherical nanoparticles," J. Chem. Phys. **120**(11), 5444-5454 (2004).
41. S. Wuestner, A. Pusch, K. L. Tsakmakidis, J. M. Hamm, and O. Hess, "Overcoming Losses with Gain in a Negative Refractive Index Metamaterial," Phys. Rev. Lett. **105**(12), 127401 (2010).
42. M. I. Tribelsky, and B. S. Luk'yanchuk, "Anomalous light scattering by small particles," Phys. Rev. Lett. **97**(26), 263902 (2006).
43. M. A. Noginov, G. Zhu, A. M. Belgrave, R. Bakker, V. M. Shalaev, E. E. Narimanov, S. Stout, E. Herz, T. Suteewong, and U. Wiesner, "Demonstration of a spaser-based nanolaser," Nature **460**(7259), 1110-1112 (2009).
44. D. Gao, A. Novitsky, T. Zhang, F. C. Cheong, L. Gao, C. T. Lim, B. Luk'yanchuk, and C.-W. Qiu, "Unveiling the correlation between non-diffracting tractor beam and its singularity in Poynting vector," Laser Photon. Rev. **9**(1), 75-82 (2015).
45. B. A. Kemp, T. M. Grzegorczyk, and J. A. Kong, "Lorentz Force on Dielectric and Magnetic Particles," J. Electromagn. Waves Appl. **20**(6), 827-839 (2006).

46. B. A. Kemp, T. M. Grzegorczyk, and J. A. Kong, "Optical Momentum Transfer to Absorbing Mie Particles," Phys. Rev. Lett. **97**(13), 133902 (2006).
47. B. Kemp, J. Kong, and T. Grzegorczyk, "Reversal of wave momentum in isotropic left-handed media," Phys. Rev. A **75**(5), 053810 (2007).
48. J. Lu, H. Yang, L. Zhou, Y. Yang, S. Luo, Q. Li, and M. Qiu, "Light-Induced Pulling and Pushing by the Synergic Effect of Optical Force and Photophoretic Force," Phys. Rev. Lett. **118**(4), 043601 (2017).

## 1. Introduction

An object is usually pushed by electromagnetic waves[1] due to the fact that the scattered radiation cannot be unidirectional. In general, the momentum transferred to the object is more than that scattered by the object in the propagation of plane waves. To obtain pulling force, one may reduce the input momentum or increase the forward scattering[1]. Employing optimized multiple plane waves with oblique incidence angle[2] one can reduce the input momentum along the z axis while maximizing the forward scattering. Instead of using multiple plane waves, Chen *et al*[3] proposed to realize optical pulling force via a Bessel light beam, which can project less photon momentum along the propagation direction and increase the forward scattering by the interference between multipoles of the particle. A proper gradientless Bessel beam can pull desired particles along both radial and longitudinal directions as a tractor beam[4-7], and recently another type of tractor beam has been demonstrated experimentally using a Gaussian beam[8]. Optical solenoid beams also possess the capability to exert a pulling force on microscopic objects[9]. Besides, light fields with phase gradients[10], optical vortices[11, 12], or nonuniform helicity[13] can be exploited for optical manipulation in alternative approaches. These proposed optical beams have promising applications in optical capture, transport and sorting various objects, such as biological molecules and living cells[14-16]. However, complex beams have their limitations in the amount of control that can be exerted. Due to unintended intensity variations[9] or short working distances[8], objects can be dragged only for several micrometers. In addition, it is still a big challenge to experimentally generate strong nonparaxial Bessel beams or uniform solenoid beams.

While it is usually considered impossible to attract particles with a uniform electromagnetic plane wave, linear momentum of light can be amplified and shows a pulling force when scattered from a lower refractive index medium into a higher one or passing through an object with gain[17, 18]. For a active system due to the momentum conservation, a tiny negative optical force is exerted on the stimulated gain medium that scatters extra momentum in the forward direction. However, it is difficult to stably manipulate particles with small optical forces because of their susceptibility to Brownian motion[19]. Enhanced optical forces were found on nanoparticles at the localized surface plasmon resonance[20, 21]. Hence, similar effects are expected at gain-assisted surface resonances and plasmon singularities. Meanwhile, plasmonic nanostructures are good candidates to sustain various plasmon resonances[22, 23]. The interaction between resonances in a nanoscale plasmonic system may give rise to Fano resonances, which were observed in both symmetric[24-27] and asymmetric[28] nanostructures. Due to their unique inherent sensitivity, Fano resonances have promising potential applications in ultrafast switches chemical sensors[29], all-optical light modulation[30] and sorting nanoparticles[31-33].

In this paper, we report enhanced optical pulling and pushing forces on an active plasmonic nanosphere around the Fano resonance. The active nanostructure is composed of an active core and a metallic shell. This type of plasmonic lasing nanostructure also features a low plasmonic lasing threshold[34]. When a uniform plane wave is incident upon the coated nanoparticle, the pulling force can be much larger than that exerted on a single active sphere[17] due to the field enhancement in the nanostructure. By tuning the radius of gain core, the subradiant quadrupole mode can be red-shifted to couple with the superradiant dipole mode, giving rise to Fano resonances. Remarkably, the pulling force can be enhanced by two orders of magnitude due to the strong field enhancement and the dominant negative Lorentz force on bound currents and charges at the Fano resonance.

## 2. Results and discussion

We consider a core-shell nanosphere with an active (gain) core and a subwavelength plasmonic (metallic) shell, surrounded by a homogeneous medium (see Fig. 1a). The relative permittivity of the metallic shell is described with the modified Drude model[35]. The gain effect of sphere core can be realized by semiconductor material or dye molecules with external pumping[36]. In this paper, we model the gain material as four level atomic system, such as fluorescent dye molecules[37, 38],

$$\varepsilon_g = \varepsilon_r + \frac{\sigma_a}{\omega^2 + i\Delta\omega_e\omega - \omega_e^2}\frac{(\tau_{21}-\tau_{10})\Gamma_{pump}}{1+(\tau_{32}+\tau_{21}+\tau_{10})\Gamma_{pump}}\frac{N_0}{\varepsilon_0} \tag{1}$$

where $\varepsilon_r = 2.05$ is the relative permittivity of the dielectric host medium, $\sigma_a = 1.71\times10^{-7}\,C^2/kg$ is the coupling strength of the polarization density to the electric field, $\tau_{ij}$ is the lifetime of the transition from state $i$ to the lower state $j$, with $\tau_{21}=3.99\,\mathrm{ns}$, $\tau_{32}=\tau_{10}=100\,\mathrm{fs}$. The center emission angular frequency is $\omega_e = 2\pi c/\lambda_e$, with angular frequency linewidth $\Delta\omega_e = 2\pi c\Delta\lambda_e/\lambda_e^2$. We assume a high concentration for the density of dye molecules as $N_0 = 8\times10^{18}\,\mathrm{cm}^{-3}$ and the pumping rate $\Gamma_{pump} = 1.5\times10^9\,\mathrm{s}^{-1}$.

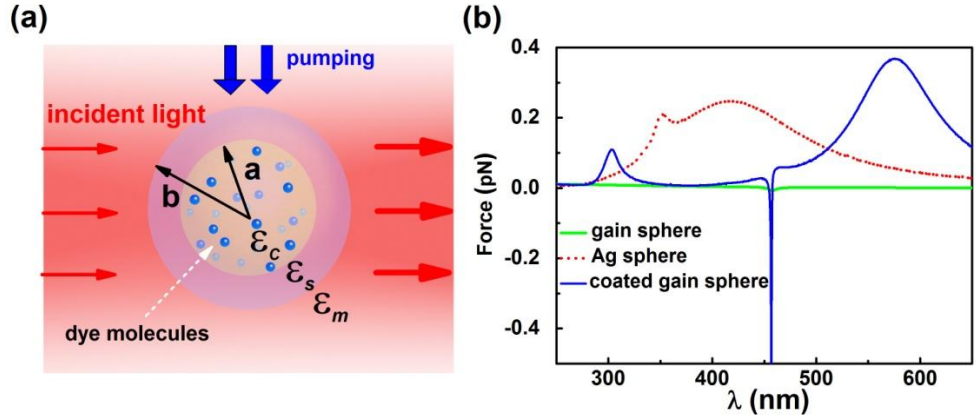

Fig. 1. (a) Schematics of the core-shell nanosphere in vacuum with plane wave incident from left. The core is made of gain material by encapsulating fluorescent dye molecules (small blue particles ) into the dielectric material. (b) The normalized forces on a single gain sphere (with radius of 60 nm), a single Ag sphere (with radius of 60 nm) and coated gain sphere with a=50 nm and b=60 nm. The pumping rate is $\Gamma_{pump} = 1.5\times10^9\,\mathrm{s}^{-1}$. In all the calculations, the intensity of incident lights is normalized to 1 mW $\mu m^{-2}$.

We calculate the optical force in the direction of wave propagation acting on the coated particle by a uniform plane wave $\mathbf{E}_{in} = E_0 e^{ikz} e^{-i\omega t}\hat{e}_x$. For simplicity, the background medium is assumed to be air, and we shall use low-power incident light so that the thermal effect on the particle can be neglected. For larger particles, higher-order eigenmodes (such as quadrupole and octupole) arise and interfere with the dipole modes[29]. Precise calculations of optical forces for large spheres can be performed by using Maxwell's stress tensor (MST),

$$\hat{T} = \frac{1}{2}(\varepsilon_0\varepsilon_m\mathbf{E}\mathbf{E}+\mu_0\mu_m\mathbf{H}\mathbf{H}) - \frac{1}{4}(\varepsilon_0\varepsilon_m|\mathbf{E}|^2+\mu_0\mu_m|\mathbf{H}|^2)\hat{\mathbf{I}}, \tag{2}$$

where $\mathbf{E}$ and $\mathbf{H}$ are the total fields obtained by Mie theory. Then the time-averaged force on the nanoparticle is the integration over any surface $\sigma$ surrounding the total coated particle[4, 5],

$$\langle \mathbf{F} \rangle = \int_{\sigma} \langle \hat{T} \cdot \mathbf{n} \rangle ds \qquad (3)$$

where $\mathbf{n}$ is the normal outwards direction to the surface $\sigma$. As shown in Fig. 1b, for coated nanospheres, two resonant peaks for the pushing forces are found at around 303 nm (bonding plasmon resonance) and around 575 nm (antibonding plasmon resonance) due to the interaction between the sphere and cavity dipole modes[39]. It should be noted that additional resonances arise from higher order modes (such as quadrupole modes at about $\lambda = 457\,\text{nm}$). The pulling force is much larger than that of on a pure dielectric gain sphere. In the following, we shall discuss the effect of active core on the pulling force in detail.

The plasmon response of this core-shell nanostructure can be easily understood by the hybridization model[40]. The hybridization between dipolar sphere and shell plasmons gives rise to a superradiant plasmon mode and a subradiant plasmon mode. Usually the quadrupole mode is suppressed due to the significant optical losses from metal shell. However, the dissipative optical loss can be compensated or even overcome by introducing gain materials into the system[41]. Once the structure becomes non-dissipative or weak dissipative, anomalous light scattering may arise around quadrupole resonance with narrow and giant lineshape[42]. The narrow quadrupole mode interfere constructively and destructively with the broad dipole mode, giving rise to Fano resonance. As is shown in Fig. 2a, the optical forces are increased with the increase of optical pumping at the Fano resonance, and even flip its direction when the pumping rate reaches the threshold $\Gamma_{\text{pump}} = 0.8 \times 10^9 \, \text{s}^{-1}$. The E-fields are greatly enhanced at the surface of the sphere and show the typical quadrupole pattern. Huge excited energies are transferred to the surface plasmon and radiate out like a 'lasing spaser'[34, 43]. In Fig. 2b, the Poynting vector lines clearly show that energy flow radiates from the gain structure. Two vortices are formed at the backward side of the particle (represented by yellow lines) due to the strong interaction of the radiated and incident waves. Around the vortices, the streamlines rotate up or down from the $xz$ plane. This counter-direction propagation cancels most of the field momentum[44] in the backward direction. In contrast, radiation in the forward direction (white lines) is emitted without the appearance of vortices, which is in favor of increasing forward linear field momentum and hence realizing large pulling forces.

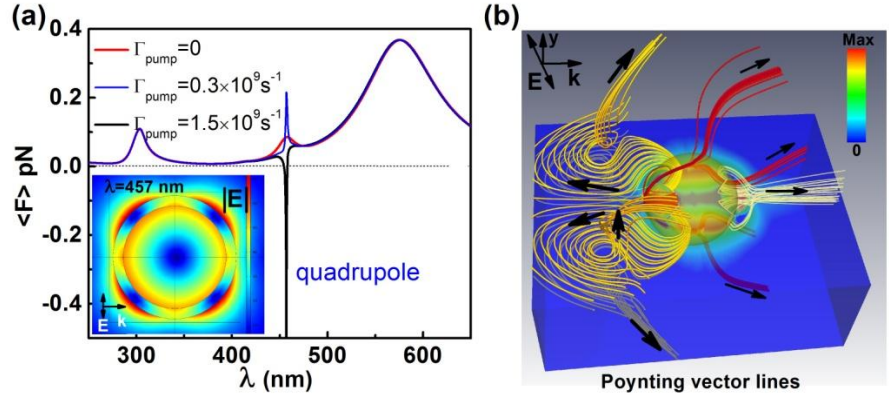

Fig. 2. (a) Optical forces on large particles with a=50 nm and b=60 nm. Insets are the electric field distributions when the particles are under pulling force. (b) 3D distributions of the Poynting vector's magnitude and Poynting vector lines when the pulling force reaches maximal at $\lambda = 457\,\text{nm}$. The streamlines demonstrate the trajectory of some representative Poynting distribution showing the emitted energy flow from the particle. The black arrows show the direction of the Poynting vectors.

Although far-field scattering of the objects could shed some information on the optical force[1], it is hard to explain quantitatively the origin of the pulling force. Besides, stimulated emission of radiation in the gain system can be overwhelmingly larger than the scattering intensity. The direction of total force has little correlation with the far-field scattering. Compared to the above MST method by the application of momentum conservation theorem, it is more fundamental to calculate the radiation pressure on an object directly using the Lorentz force[45]. The momentum contributions can be separated into forces on bound currents and charges ($\mathbf{F}_b$) and on free currents ($\mathbf{F}_c$)[46]. In this paper, we only consider nonmagnetic materials, so the bound current density from the magnetization $\nabla \times \mathbf{M}$ and the magnetic charge density $-\nabla \cdot \mathbf{M}$ are zero. Then the two parts of the Lorentz force density $\mathbf{f}_b$ and $\mathbf{f}_c$ can be reduced as follows [46],

$$\mathbf{f}_b = \frac{1}{2}\text{Re}\left[\varepsilon_0 (\nabla \cdot \mathbf{E})\mathbf{E}^* - i\omega\varepsilon_0 (\varepsilon_R - 1)\mathbf{E} \times \mathbf{B}^*\right]$$
$$\mathbf{f}_c = \frac{1}{2}\text{Re}\left[\omega\varepsilon_I \mathbf{E} \times \mathbf{B}^*\right] \qquad (4)$$

where * represents the complex conjugate of a quantity and $\varepsilon_R (\varepsilon_I)$ is the real (imaginary) part of the relative permittivity in the core or shell. The first and second terms of $\mathbf{f}_b$ are the force contributions from bound electric charges and bound electric currents, respectively. The total Lorentz force on the core-shell nanostructure is the sum of the integration of $\mathbf{f}_b$ and $\mathbf{f}_c$ over the whole structure. In the following, we use the finite element method (COMSOL MULTIPHYSIS V.5.0) to obtain the Lorentz force and verify the results from Mie theory by MST method.

In a lossy system, the force density on the bound electric charges and currents $\mathbf{f}_b$ can be positive or negative depending on the polarization $\mathbf{P}$ at that position. The force density on free currents $\mathbf{f}_c$ is a pushing force because the incident light attenuates and transfers momentum to free currents in the lossy media[47]. However, in a gain medium the light is magnified and produces extra momentum to the free currents, resulting in negative $\mathbf{f}_c$. When the magnitude of $\mathbf{F}_c$ is larger than $\mathbf{F}_b$, the total force will become a pulling force and drag the particle towards the light source. This also explains why there is a threshold gain for a gain sphere to be pulled by a plane wave[17]. If the particle has the same real part of the permittivity as the background media, there will be no bound electric charge or current at the boundary. Then $f_c$ disappears and the threshold gain vanishes as well.

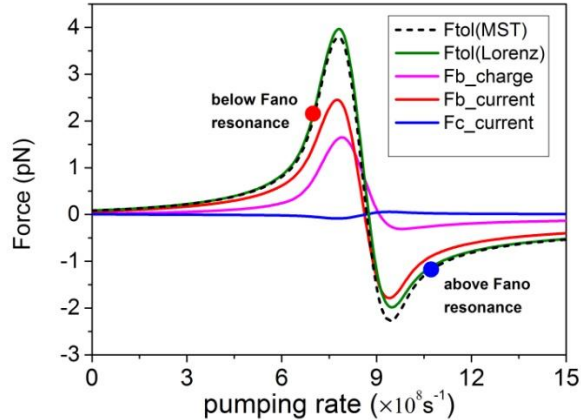

Fig. 3. Forces on a core-shell nanostructure (a = 50 nm and b=60 nm) for different pumping rate. The total Lorentz forces are divided into two parts: $\mathbf{F}_b$ and $\mathbf{F}_c$, respectively. The results from Lorentz force show good consistency with the results from the MST method. The observable divergence may due to the vast spatial variation of Lorentz force density[45] when the field intensity are greatly enhanced.

Core-shell nano-structures have the advantage of low-threshold pumping in realizing plasmon stimulated emissions. Dipole surface plasmon modes in the metallic shell can be excited in this core-shell nanostructure enhancing the electric field in the gain core[34]. However, with the increase of pumping rate, the metallic losses from shell are compensated around the quadrupole resonance, making the strength of quadrupole mode comparable to that of the dipole mode. Then the narrow dark quadrupole mode will couple with the broad bright dipole mode, giving rise to a Fano resonance. As is shown in Fig. 3, the induced bound force $\mathbf{F}_b$ is dominant, while the force on free currents is basically negligible. When the pumping rate is increased, the field intensity is greatly increased around Fano resonance. However, the enhancement of optical force does not come from the force on free currents in the gain core (the force on free currents of the core remains the same), but from the large force $f_b$ on the metal shell. This is easy to understand: the bound current density $\mathbf{J}_b = \partial \mathbf{P}/\partial t = -i\omega\varepsilon_0(\varepsilon-1)\mathbf{E}$ is increased with the enhanced E-field. Large $\mathbf{f}_b$ corresponds to a large local polarization density.

Further increasing the pumping rate, the quadrupole mode and dipole mode will interact destructively above Fano resonance. The maximal pulling force is about two orders larger than that of a single gain particle. More interestingly, the optical force on core-shell particles will not only be greatly enhanced, but also reverse its direction around Fano resonance. By tuning the pumping rate, one could switch the pulling force and pushing force on active plasmonic nanoparticles. Similar to the pulling-pushing effect on a tapered fiber by photophoretic force[48], external pumping may provide an alternative way to transport nanoparticles back and forth. The interference of two radiative modes generates a very complex near field distribution, presenting optical vortices of energy flow $\mathbf{E}\times\mathbf{H}^*$ inside the nanoparticle[42]. The handedness of the optical vortex is reversed due to the change of interaction ways from constructive to destructive. From Eq. (4), one can see that the magnitude and direction of energy flow can alter the force density on induced currents, and therefore change the direction of total force. Fig. 4 shows the induce charge distribution and current distribution below Fano resonance and above Fano resonance, respectively. Both induced bound charges at the core-shell surfaces and induced current inside the particle flip around Fano resonance, leading to the reversal of $\mathbf{f}_b$.

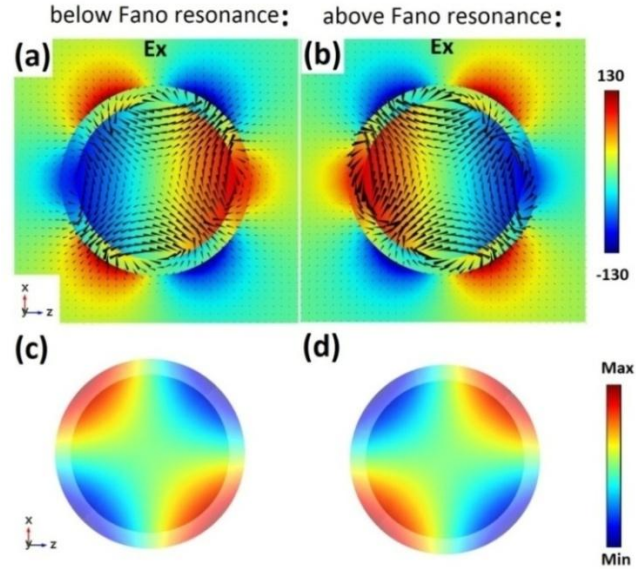

Fig. 4. The distributions of Ex (a, b) and charge density (c, d) at the surfaces of core-shell particles around Fano resonances. The below Fano resonances ($\Gamma_{pump}=0.7\times10^9\,\mathrm{s}^{-1}$) and above Fano resonances ($\Gamma_{pump}=1.1\times10^9\,\mathrm{s}^{-1}$) are corresponding to the red and blue dot in Fig. 3. The arrows represent the direction of induced current.

## 3. Conclusions

In conclusion, large optical forces are demonstrated on plasmonic nanostructure particles with gain. Pulling and pushing optical force can be switched in different plasmonic modes by controlling the incident wavelength. Large enhancements of negative optical force are achieved around Fano resonance. The giant negative force mainly roots in the Lorentz force on bound currents and charges of the metallic shell. Our work may give deep insight into the mechanism of pulling force in gain systems and offer an effective way to obtain large negative forces for nano manipulation.

**Funding**

National Natural Science Foundation of China (Grant No. 11374223, No.11504252), National Science of Jiangsu Province (Grant No. BK20161210), Natural Science Foundation for the Youth of Jiangsu Province(No. BK20150306), Qing Lan project, "333" project (Grant No. BRA2015353), Natural Science Foundation for Colleges and Universities in Jiangsu Province of China (No.15KJB140008) and PAPD of Jiangsu Higher Education Institutions.